\documentclass{arxiv}

\usepackage{graphicx}
\usepackage{amsmath}
\usepackage{amsfonts}
\usepackage{amssymb}

\usepackage[symbol]{footmisc}

\usepackage{url}

\usepackage{graphicx}
\usepackage{epstopdf}

\usepackage{tikz}
\usetikzlibrary{decorations.markings}
\usetikzlibrary{shapes,snakes}
\usetikzlibrary{shapes.geometric}
\usetikzlibrary{fit}                                    
\usetikzlibrary{backgrounds}
\usetikzlibrary{positioning}

\tikzstyle{every picture}+=[remember picture]
\tikzstyle{na} = [baseline=-.5ex]

\usepackage{listings}
\definecolor{fdbbColor}{RGB}{0,166,214}
\definecolor{uetliColor}{RGB}{230,70,22}
\definecolor{etlColor}{RGB}{107,134,137}

\title{FDBB: Fluid Dynamics Building Blocks}

\author{Matthias M¨\"oller$^{1,}$\footnote{Corresponding author: \url{m.moller@tudelft.nl}}, Andrzej Jaeschke$^{2}$}

\heading{Matthias M\"oller, Andrzej Jaeschke}

\address{
$^{1}$ Delft University of Technology, 
Delft Institute of Applied Mathematics,\\
Van Mourik Broekmanweg 6,
2628 XE Delft, The Netherlands\\
$^{2}$ \L{}\'od\'z University of Technology,
Institute of Turbomachinery,\\
ul. W\'olcza\'nska 219/223, 90-924 \L{}\'od\'z, Poland,
}


\abstract{
High-performance computing platforms are becoming more and more heterogeneous, which makes it very difficult for researchers and scientific software developers to keep up with the rapid changes on the hardware market. In this paper, the open-source project FDBB (Fluid Dynamics Building Blocks) is presented, which eases the development of fluid dynamics applications for heterogeneous systems. It consists of a low-level API that provides a unified interface to many different linear algebra back-ends and a lightweight and extendible high-level expression template library, which provides largely customizable fluid dynamics building blocks, like transformations between primary and secondary variables as well as expressions for Riemann invariants, equations of state, inviscid fluxes and their flux-Jacobians. The performance of the developed approach is assessed both for synthetic micro-benchmarks and within mini-applications.
}

\begin{document}

\thispagestyle{empty}

\section{INTRODUCTION}
\label{sec:introduction}

High-performance computing hardware is progressing quite rapidly towards more and more heterogeneous platforms, which makes it very difficult for researchers and developers of scientific software to keep up with the latest developments in chip designs and to explore emerging hardware technologies like, e.g., hybrid CPU-FPGA devices, without spending a large part of their time on reimplementing the same algorithms and core functionalities over and over again for the different compute devices. Software packages that explicitly aim at supporting heterogeneous platforms, e.g., multi-core CPU systems combined with many-core accelerators like GPUs, dedicated vector processors, and/or FPGA expansion cards, suffer even more from the rapid changes of the hardware market, since their developers need to ensure that the implementations of an algorithm for the different hardware platforms are kept at the same maturity and functionality level.

Over the last two decades, the trend in high-performance computing (HPC) for practical large-scale applications goes away from writing hand-optimized application codes towards compiler-based code generation and automated performance tuning \cite{XTune,Patus}. 

It is a long-living myth that the use of expression template meta-programming techniques \emph{automagically} leads to efficient computer code. However, the use of \emph{specialized} expression template libraries (ETL) like Armadillo \cite{Armadillo}, ArrayFire \cite{ArrayFire}, Blaze \cite{Blaze}, Eigen \cite{Eigen}, VexCL \cite{VexCL}, ViennaCL \cite{ViennaCL} that provide hardware-optimized linear algebra routines for vectors as well as dense and sparse matrices allows to write concise source code at an abstract mathematical level that gets compiled into executables that exploit the hardware capabilities to the extent implemented by the authors of the linear algebra back-ends. None of them, of course, supports all target hardware platforms and the provided functionality and foreseen use case scenarios largely differ from one library to the other. Despite all differences, the common aim of expression template libraries is to provide mechanisms to formulate mathematical vector expressions like \lstinline{y=0.5*sin(x+y)} and evaluate them in a single loop over the vector entries rather than creating temporaries for sub-expression.

The Fluid Dynamic Building Blocks (FDBB) project \cite{FDBB} makes an attempt to develop a unified wrapper interface for the core functionality of \emph{all} of the aforementioned linear algebra back-ends and provides an extendible set of expression templates for developing fluid dynamic applications with the focus placed on compressible flows. These expressions include transformations between conservative, primitive and characteristic variables as well as Riemann invariants, different types of equations of state (EOS), as well as inviscid fluxes and their flux Jacobians. FDBB is a header-only C++11/14 library, which is designed to leave the underlying linear algebra back-ends largely unmodified so that applications automatically benefit from improvements in the ETLs provided that their API does not change between versions. The low-order API of our package is released as standalone software, the Unified Expression Template Library Interface (UETLI) \cite{UETLI}.

The rest of the paper is structured as follows. Section~\ref{sec:implementation} describes the implementation and typical usage scenarios in more detail. A performance analysis of the low- and high-level APIs is presented in Section~\ref{sec:performance_analysis} followed by conclusions drawn in Section~\ref{sec:conclusions}.

\begin{figure}[b!]
\begin{center}
\begin{tikzpicture}[scale=0.95,font=\sffamily\small]
\draw [etlColor, very thick] (-1.1,0) rectangle (-0.1,2) node[etlColor, pos=0.5, rotate=90] {ETLs};
\draw [etlColor, very thick] (0,0) rectangle (1,2) node[pos=0.5, rotate=90] {Armadillo};
\draw [etlColor, very thick] (1,0) rectangle (2,2) node[pos=0.5, rotate=90] {ArrayFire};
\draw [etlColor, very thick] (2,0) rectangle (3,2) node[pos=0.5, rotate=90] {Blaze};
\draw [etlColor, very thick] (3,0) rectangle (4,2) node[pos=0.5, rotate=90] {Blitz++};
\draw [etlColor, very thick] (4,0) rectangle (5,2) node[pos=0.5, rotate=90] {Eigen};
\draw [etlColor, very thick] (5,0) rectangle (6,2) node[pos=0.5, rotate=90] {IT++};
\draw [etlColor, very thick] (6,0) rectangle (7,2) node[pos=0.5, rotate=90] {MTL4};
\draw [etlColor, very thick] (7,0) rectangle (8,2) node[pos=0.5, rotate=90] {uBLAS};
\draw [etlColor, very thick] (8,0) rectangle (9,2) node[pos=0.5, rotate=90] {VexCL};
\draw [etlColor, very thick] (9,0) rectangle (10,2) node[pos=0.5, rotate=90] {ViennaCL};
\draw [etlColor, very thick] (10,0) rectangle (11,2) node[pos=0.5, rotate=90] {...};
\draw [uetliColor, very thick, fill=uetliColor] (-1.1,2.1) rectangle (-0.1,4) node[white, pos=0.5, rotate=90] {UETLI};
\draw [uetliColor, very thick, fill=uetliColor] (0,2.1) rectangle (11,4) node[white, pos=0.5, text width=10cm] {Unified function wrapper API to core functionality of ETL's: make\_temp, tag, tie, arithmetic operations, caching, ...};
\draw [fdbbColor, very thick, fill=fdbbColor] (-1.1,4.1) rectangle (-0.1,6) node[white, pos=0.5, rotate=90] {FDBB};
\draw [fdbbColor, very thick, fill=fdbbColor] (0,4.1) rectangle (11,6) node[white, pos=0.5, text width=10cm] {ETs for conservative/primitive/characteristic variables, EOS, inviscid/viscous fluxes, flux-Jacobians,  Riemann invariants};
\end{tikzpicture}
\caption{Structure of the low-level Unified Expression Template Library Interface (UETLI) and the high-level Fluid Dynamics Building Blocks (FDBB) open-source header-only C++11 library.}
\label{fig:FDBB}
\end{center}
\end{figure}

\section{IMPLEMENTATION}
\label{sec:implementation}

The overall structure of our software package is depicted in Figure~\ref{fig:FDBB}.

\subsection{Low-level API: UETLI}
\label{sec:implementation_uetli}

\paragraph{Core Functionality}

The different linear algebra back-ends largely vary in functionality, maturity, performance, calling conventions and in the way they evaluate the expressions on the target hardware. Most CPU back-ends employ a delayed evaluation approach based on recursive templates and template meta-programming, to combine several operations into one to reduce (or eliminate) the need for temporaries. In contrast, the multi-device back-ends ArrayFire \cite{ArrayFire}, VexCL \cite{VexCL} and ViennaCL \cite{ViennaCL} utilize just-in-time (JIT) compilation techniques to convert automatically generated source code into executable code. 

\begin{figure}[h!]
\begin{lstlisting}
vex::vector<float> x ( ctx, n );
vex::vector<float> y ( ctx, n );

fdbb::tag<1>(y) = CONSTANT( 0.5, y ) 
                * fdbb::elem_sin( fdbb::tag<0>(x)+fdbb::tag<1>(y) );
\end{lstlisting}
\caption{Code snippet for the evaluation of the vector expression $y=0.5\sin(x+y)$.}
\label{fig:code_expression1}
\end{figure}

Consider the code snippet depicted in Figure~\ref{fig:code_expression1} that computes the element-wise sine of the vector sum $x+y$, scales the result by the constant 0.5 and assigns it to $y$. The \lstinline{tag<ID>(expression)} function is optional to assign a unique ID tag to the expression, which helps the VexCL \cite{VexCL} back-end to not pass the same expression as multiple arguments to the device kernel. As a general design principle of our software, functionality that is only supported by some ETLs reduce to no-ops in the other cases, which is realized by template specialization. The \lstinline{CONSANT(value, expression)} macro ensures that the data type of the constant equals that of the expression result and, moreover, enables further optimization if that is provided by the back-end. The unitary operation \lstinline{elem_sin(expression)} is one example of more than 30 element-wise arithmetic operations that can be applied to vectors, dense and sparse matrices, and block expressions, the latter being discussed below.

\begin{figure}[h!]
\begin{lstlisting}
kernel void vexcl_vector_kernel ( ulong n,
                                  global float * prm_tag_1_1,
                                  global float * prm_tag_0_1 )
{
  for(ulong idx = get_global_id(0); idx < n; idx += get_global_size(0))
  {
    prm_tag_1_1[idx] = ( ( 5.0000000000000000e-01f ) * 
      sin( ( prm_tag_0_1[idx] + prm_tag_1_1[idx] ) ) );
  }
}
\end{lstlisting}
\caption{Auto-generated OpenCL kernel code for the expression $y=0.5\sin(x+y)$.}
\label{fig:code_kernel1}
\end{figure}

Figure~\ref{fig:code_kernel1} shows the OpenCL source-code that has been auto-generated by the VexCL \cite{VexCL} library from the above vector expression and can be further processed into CPU or GPU code by the OpenCL subsystem. VexCL also provides code generation engines for CUDA and OpenMP as well as 
experimental support for Maxeler's dataflow computing platform \cite{MaxJ}, which aims at making field-programmable gate arrays (FPGAs) usable as next-generation accelerator devices. Maxeler Technologies provides a software development kit consisting of a Java-like programming language, MAXJ, as well as compilers and libraries to synthesizes the high-level compute kernels into bitstreams to reconfigure the FPGAs at runtime. In this case, JIT compilation can take up to several hours or days but, still, the fully automated generation of FPGA bitstreams from mathematical expressions is an attractive feature. Switching to another back-end is realized by changing lines 1--2 of Figure~\ref{fig:code_expression1} leaving the actual mathematical expression in lines 3-4 unmodified.

\paragraph{Caching Mechanism.}

The library makes extensive use of rvalue references, move semantics, and perfect forwarding. Wrapper functions are implemented based on the design pattern depicted in Figure~\ref{fig:code_move_semantics}. 
If type \lstinline{A} requires special treatment in back-end \lstinline{BACKEND} other than calling \lstinline{sin(std::forward<A>(a))}, the \lstinline{get_element_sin_impl<A>} trait needs to be specialized to hold  \lstinline{EnumETL::BACKEND} in its attribute \lstinline{value}. The functionality is then provided by the specialized function \lstinline{elem_sin_impl<A,EnumETL::BACKEND>::eval(A&& a)}. 

\begin{figure}[h!]
\begin{lstlisting}
template<typename A>
auto elem_sin(A&& a)
     #if __cplusplus <= 201103L
     -> decltype(...) // C++11 return type deduction
     #endif
{
  return backend::detail::elem_sin_impl<A,
         backend::detail::get_elem_sin_impl<A>::value>
         ::eval(std::forward<A>(a));
}
\end{lstlisting}
\caption{Static back-end dispatching: Example implementation of the element-wise sine expression.}
\label{fig:code_move_semantics}
\end{figure} 

This approach enables minimally-invasive addition of back-ends and fine-grained control over specialized treatment at the level argument types in unary and binary operators.

The \lstinline{elem_sin(a)} function automatically returns an expression object, whose type depends on the adopted linear algebra back-end. It can be either assigned (=evaluated) to a vector or matrix object, i.e. \lstinline{y = elem_sin(x)}, or passed as argument to another function, i.e. \lstinline{y = elem_sin(elem_sin(x))}. The latter is used on Section~\ref{sec:implementation_fdbb} to compose expressions for the inviscid fluxes and the flux-Jacobians from smaller modular building blocks.

However, not all back-ends support the construction of expressions from sub-expressions, which is caused by their inability to store temporarily created sub-expressions as \emph{objects} in further expressions but instead just store \emph{references}, which will become invalid once the underlying objects reach the end of their scope. We have implemented a caching mechanism as a remedy, which is itself a lightweight ETL with full UELTI functionality support. Cache expressions encapsulate the temporal expression \emph{objects} that are generated by the back-end and pass references to them transparently to the back-end functions. 

\begin{figure}[h!]
\begin{lstlisting}
auto x = fdbb::cache::CacheExpr<0, arma::vec>( arma::vec(10) ); 
auto y = fdbb::cache::CacheExpr<1, arma::vec>( arma::vec(10) ); 

auto E = CONSTANT( 0.5, y) * fdbb::elem_sin( x+y );
     y = E;
\end{lstlisting}
\caption{Code snippet illustrating the caching mechanism.}
\label{fig:code_caching}
\end{figure} 

Figure~\ref{fig:code_caching} illustrates the general use of the caching mechanism for the Armadillo \cite{Armadillo} back-end. The two column vectors  are encapsulated by the \lstinline{CacheExpr<Tag,ExprType>} objects, which require unique ID tags next to the expression types. All unary and binary operations return themselves cache-type objects that hold the sub-expression objects internally. The expression object can be obtained using the \lstinline{get()} function, which makes it even possible to combine the caching mechanism with native expression template code
\begin{center}
\lstinline{x.get() = y.get() + arma::randu(10);}
\end{center}
It should be noted that line 5 in the above code snippet does not trigger evaluation of the cached expression but only assigns the unevaluated and encapsulated expression to the object \lstinline{E}. Evaluation happens during the assignment to \lstinline{y} in line 6. In practice, lines 5-6 are typically fused unless \lstinline{E} serves as sub-expression in multiple further expressions.

The caching mechanisms is moreover a helpful debugging tool. \lstinline{E.pretty_print(os)} yields
\begin{center}
\lstinline{(E(0.5)*sin((E(N4arma3ColIdEE)+E(N4arma3ColIdEE))))}
\end{center}
An extensive dump of the entire expression tree is produced by \lstinline{E.print_debug(os)}.

\paragraph{Block Expressions.}

Lastly, UETLI provides a framework for working with block expressions, that is, expressions that are composed from block matrices and vectors of fixed block size. As an example, consider the following block matrix-vector multiplication
\begin{equation}
\begin{bmatrix}
y_0\\
y_1
\end{bmatrix}
=
\begin{bmatrix}
A & B\\
C & D
\end{bmatrix}
\begin{bmatrix}
\sin(x_0+x_1)\\
\cos(x_0-x_1)
\end{bmatrix},
\label{eq:expression2}
\end{equation}
where $x_0,x_1\in\mathbb{R}^n$, $y_0,y_1\in\mathbb{R}^m$, and $A,B,C,D\in\mathbb{R}^{m\times n}$. With the aid of block expressions, this can be implemented as shown in the code snippet depicted in Figure~\ref{fig:code_expression2}. Here, \lstinline{M} and \lstinline{Y} (lines 9-12) are a block matrix \emph{view} and block column vector, respectively, and \lstinline{E} (lines 13-14) is a block expression consisting of 2 rows and 1 column. Block matrices and vectors can only store objects of the same type, whereas block expressions accept an arbitrary combination of types as it is required to handle the different expression objects returned from the element-wise sine and cosine function. View objects in contrast to the non-view counterparts only store references to the arguments passed, which implies that the sparse scalar matrices \lstinline{A}, \lstinline{B}, \lstinline{C}, and \lstinline{D} are not duplicated and copied into \lstinline{M} (line 9), which would have been the case if \lstinline{M} was of \lstinline{BlockMatrix} type. In contrast, the move constructor of the block column vector is used in lines 10-12 so that, again, no data duplication takes place.

\begin{figure}[t!]
\begin{lstlisting}
vex::vector<double> x0 ( ctx, n );
vex::vector<double> x1 ( ctx, n );
// Initialize row, col, and value vectors for matrices beforehand...
vex::sparse::matrix<float> A ( ctx, m, n, rowA, colA, valA );
vex::sparse::matrix<float> B ( ctx, m, n, rowB, colB, valB );
vex::sparse::matrix<float> C ( ctx, m, n, rowC, colC, valC );
vex::sparse::matrix<float> D ( ctx, m, n, rowD, colD, valD );

fdbb::BlockMatrixView<vex::sparse::matrix<float>,2,2> M ( A, B, C, D );
fdbb::BlockColVector<vex::vector<double>,2>           Y ( 
                                      vex::vector<double> ( ctx, m ), 
	                                  vex::vector<double> ( ctx, m ) );                                
auto E  = fdbb::makeBlockExpr<2,1> ( fdbb::elem_sin( x0+x1 ),
                                     fdbb::elem_cos( x0-x1 ) );
Y = M * E;

\end{lstlisting}
\caption{Code snippet for the block matrix-vector multiplication in Eq. \eqref{eq:expression2}.}
\label{fig:code_expression2}
\end{figure}

\lstinline{makeBlockExpr(exprs...)} creates an object of type \lstinline{BlockExpr<nrow,ncol,Exprs...>}, which is the most flexible block container since it supports the mixing of back-ends by design. However, most linear-algebra back-ends do not support mixed expressions, i.e., matrix-vector multiplication between a sparse Eigen matrix and a Blaze vector (yet). The aforementioned block types support all unitary and binary operations and element-wise functions that can be applied to a scalar matrix or vector object if they make mathematical sense.

The \lstinline{idx}-th sub-item of a block object is accessible via \lstinline{fdbb::utils::get<idx>(obj)}, whereby all block types adopt row-major storage ordering by default. The latter can be adjusted by passing \lstinline{StorageOrder::ColMajor} as template parameter to the block objects.

\subsection{High-level API: FDBB}
\label{sec:implementation_fdbb}

On top of UETLI, we created the expression template library FDBB, which provides the main building blocks for developing fluid dynamics applications. 

\paragraph{Variables.}
Secondary variables can be computed from the primary ones with only a few lines of code. Let $U=[\rho,\rho\mathbf{v},\rho E]^\top$ denote the state vector of conservative variables in 3D and assume a perfect gas with adiabatic index $\gamma=c_p/c_v=1.4$. Then the absolute pressure $p$ is computed in a single line as shown in the code snippet depicted in Figure~\ref{fig:code_variables1}. By making dimension $N_D$ (\lstinline{'3'}) and variable type (\lstinline{'EnumForm::conservative'}) template parameters, it is even possible to write dimension- and formulation-independent application code. This approach is most effective in combination with factory-based object creation \cite{Gamma}.

\begin{figure}[h!]
\begin{lstlisting}
using  eos = fdbb::EOSidealGas<double, 
                               std::ratio<7, 2> /* Cp */,
                               std::ratio<5, 2> /* Cv */>;
using varU = fdbb::Variables<eos,3,fdbb::EnumForm::conservative>;
// Create and fill vectors rho, mx, my, mz, and rhoE beforehand...
auto     p = varU::p( rho, mx, my, mz, rhoE );
\end{lstlisting}
\caption{Code snippet for computing the pressure $p$ from the conservative variables $U=[\rho,\rho\mathbf{v},\rho E]^\top$.}
\label{fig:code_variables1}
\end{figure}

Instead of passing the scalar variables one by one, it is handy to collect them in a block expression that can be passed as single parameter (see paragraph 'Passing of Variables' below for details). The scalar variables can be accessed via \lstinline{fdbb::utils::get<idx>(U)}.

\begin{figure}[h	!]
\begin{lstlisting}
auto U = varU::conservative ( rho, mx, my, mz, rhoE );
auto p = varU::p( U );
\end{lstlisting}
\caption{Collection of state variables into block expressions}
\label{fig:code_variables2}
\end{figure}

A further advantage of using block expression is the easy conversion between state vectors. Let the vector of conservative values \lstinline{U} be defined as in line 1 of Figure~\ref{fig:code_variables2}. Then the conversion from conservative to primitive variables and vice versa can be realized elegantly as illustrated in the following code snippet. Assuming that well-designed linear algebra back-ends will not perform copy operations if source and destination vectors are the same, no memory bandwidth is lost on unnecessary transfers of the density variable. 

\begin{figure}[h	!]
\begin{lstlisting}
using varV = fdbb::Variables<eos,3,fdbb::EnumForm::primitive>;
auto     V = varV::conservative ( rho, vx, vy, vz, p );

         V = varU::primitive   ( U );
         U = varV::conservative( V );
\end{lstlisting}
\caption{Conversion from conservative to primitive variables and vice versa.}
\label{fig:code_variables3}
\end{figure}

\paragraph{Equations of state.}

User-defined equations of state of the form $f(p,V,T)=0$ with absolute pressure $p$, volume $V$, and absolute temperature $T$ can be specified by implementing a derived class that implements the prototype \lstinline{EOF_pVT} depicted in Figure~\ref{fig:code_eos_prototype}. In a forthcoming release of FDBB, experimental support for the open-source thermophysical property library CoolProp \cite{CoolProp} will be enabled, which provides a large collection of equations of state for (pseudo-)pure fluids. Since CoolProp does not accept vector expressions as arguments, its use is currently limited due to the generation of temporary objects. This shortcoming can be overcome by extending CoolProp to accept generic vector arguments and perform computations based on expression templates rather than data directly.

\begin{figure}[h	!]
\begin{lstlisting}
struct userDefinedEOS : public EOS_pVT
{
  template<typename Trho, typename Te>
  static FDBB_INLINE auto constexpr p_rhoe(Trho&& rho, Te&& e);

  template<typename Trho, typename Te>
  static FDBB_INLINE auto constexpr T_rhoe(Trho&& rho, Te&& e);
  ...  
  static std::ostream& print(std::ostream& os);
};
\end{lstlisting}
\caption{Prototype of a user-defined equation of state of the form $f(p,V,T)=0$.}
\label{fig:code_eos_prototype}
\end{figure}

\paragraph{Inviscid Fluxes.} The $N_U\times N_D$ dimensional tensor of inviscid fluxes
$$
\mathbf{F}(U)=
\begin{bmatrix}
\rho\mathbf{v}\\
\rho\mathbf{v}\otimes\mathbf{v}+\mathcal{I}p\\
\mathbf{v}(\rho E+p)
\end{bmatrix}
$$
is implemented as a ready-to-use block expression, cf. Figure~\ref{fig:code_inviscid_fluxes1}, whereby it is assumed that \lstinline{eos} and \lstinline{varU} are defined as in lines 1--4 of Figure~\ref{fig:code_variables1} and \lstinline{U} is the state vector of conservative variables defined in line 1 of Figure~\ref{fig:code_variables2}. It should be noted that \lstinline{F} only holds the expressions, while their evaluation takes place upon assignment to block matrix \lstinline{f}.
\begin{figure}[h!]
\begin{lstlisting}
using fluxU = fdbb::Fluxes<varU>;
auto      F = fluxU::inviscid ( U );
fdbb::BlockMatrix<vec_t,5,3> f( F );
\end{lstlisting}
\caption{Code snippet for computing the inviscid fluxes for conservative state variables.}
\label{fig:code_inviscid_fluxes1}
\end{figure}

The implementation of flux-Jacobian matrices for the inviscid fluxes is not yet finished and will be enabled in a forthcoming release of the FDBB library.

\paragraph{Passing of Variables.} 

FDBB has been designed with utmost flexibility in mind. Except for a few exceptions, all functions exhibit the same generic interface shown in Figure~\ref{fig:code_variable_passing1},
\begin{figure}[b!]
\begin{lstlisting}
template<typename... Vars>
static auto constexpr conservative(Vars&&... vars)->decltype(...){...}
\end{lstlisting}
\caption{Code snippet for computing the inviscid fluxes for conservative state variables.}
\label{fig:code_variable_passing1}
\end{figure}
which allows the user to pass any combination of arguments and leave the mapping to variables to an extra trait that is passed to the variable type, say \lstinline{varU}, as additional template parameter. The default behavior is a perfectly forwarding 1-to-1 map from the parameter pack \lstinline{vars...} to the variables, whereby the mapping from the variable, dimension, and formulation triple to the argument index is realized via an extensive specialization of the \lstinline{MapVar2Arg<Var, dim, Form>} trait. The default behavior can be changed by providing a user-defined mapping that follows the structure depicted in Figure~\ref{fig:code_variable_passing2}. Additional traits the support the passing of state variables as block objects as illustrated in Figures~\ref{fig:code_variables2} and \ref{fig:code_variables3} are provided and can be further adjusted to the needs of the user.
\begin{figure}[h!]
\begin{lstlisting}
template<std::size_t dim, fdbb::EnumForm Form>
struct TraitsPerfectForwarding
{
  template<fdbb::EnumVar var, typename... Vars>
  static auto constexpr getVariable(Vars&&... vars) noexcept
    -> const
    typename std::tuple_element<fdbb::MapVar2Arg<var, dim, Form>::index,
                                std::tuple<Vars...>>::type
  {
    return std::get<fdbb::MapVar2Arg<var, dim, Form>::index>(
      std::tuple<Vars...>(vars...));
  }
};
\end{lstlisting}
\caption{Code snippet for a perfectly forwarding 1-to-1 map from arguments to variables.}
\label{fig:code_variable_passing2}
\end{figure}

\section{PERFORMANCE ANALYSIS}
\label{sec:performance_analysis}

An extensive performance analysis of all possible combinations of linear algebra back-ends and FDBB features on all supported hardware platforms is beyond the scope of this paper. We restrict ourselves to a synthetic micro-benchmark to measure the computational overhead introduced by the extra FDBB layer and one mini-application.

\paragraph{Micro-benchmark.} The kinetic energy or a multiple thereof occurs quite frequently as sub-expression in fluid dynamics applications. We therefore chose the calculation of $\|\mathbf{v}\|^2$ from the conservative state vector in 3D, i.e. \lstinline{varU::v_mag2(U)}, as micro-benchmark. 

All tests were run under CentOS Linux 6.7 with thread pinning (\lstinline{likwid-pin -c N:0-15}) on a dual-socket workstation (Intel E5-2670 @ 2.6 GHz, 20MB cache) with 64GB main memory. The ArrayFire and VexCL back-ends were tested in CUDA-mode on an NVIDIA Tesla K20Xm GPU with 6GB memory and ECC turned off. The exact compiler versions were gcc 5.3.0 and nvcc 7.5.17 with CUDA driver version 352.93.

\begin{figure}[h!]
\begin{center}
\resizebox{\textwidth}{!}{
\sffamily
\input{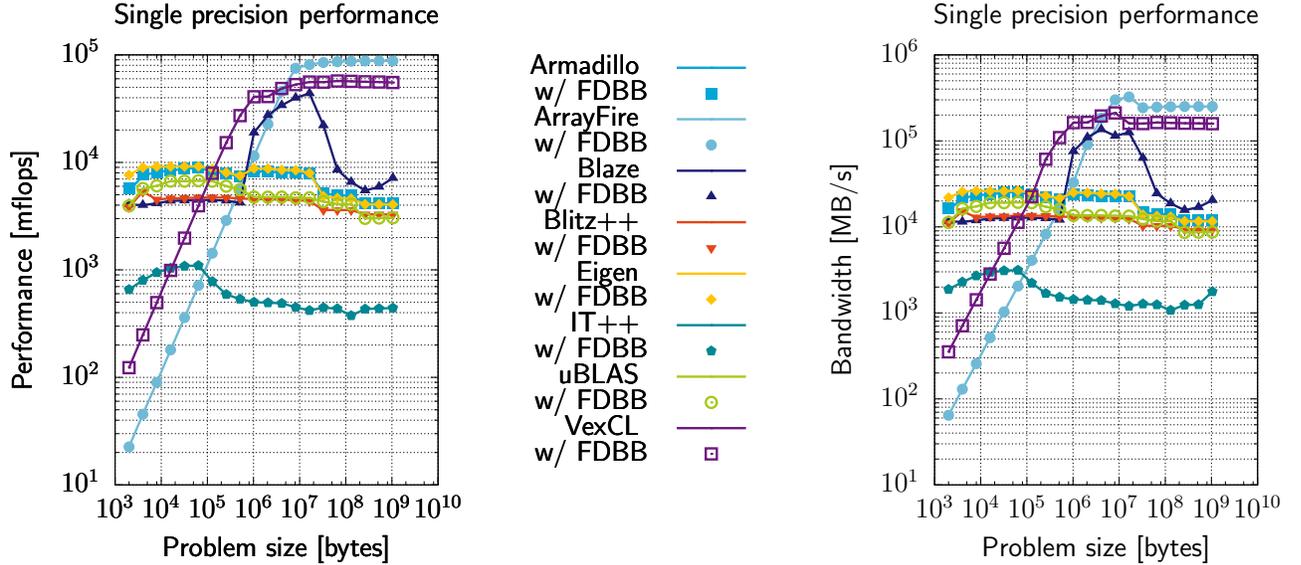}}
\end{center}
\caption{Compute performance (left) and memory bandwidth (right) for the expression given in Eq. \eqref{eq:micro-benchmark} computed in single precision with the different linear algebra back-ends on CPUs and GPUs.}
\label{fig:test1_sp}
\end{figure}

Figure~\ref{fig:test1_sp} shows the compute performance (left) and the memory bandwidth (right) measured for a wide range of problem sizes for the element-wise  expression
\begin{equation}
y\leftarrow (m_x.*m_x+m_y.*m_y+m_z.*m_z)./(\rho.*\rho).
\label{eq:micro-benchmark}
\end{equation}
Remarkably, no measurable performance loss is observed between the back-end specific implementations (straight lines) and the FDBB-enabled generic ones (symbols).

\paragraph{Mini-application.} To estimate the performance of FDBB in real-life scenarios we implemented a mini-app, in which the conservative variables are initialized once by physical values and used to evaluate the inviscid fluxes multiple times. The computing times are given in Table~\ref{tab:mini-app}. Columns 2--4 show the wall clock-times (in $\mu s$) measured for the three efficient linear algebra back-ends Blaze, Eigen, and VexCL. Though all back-ends run in CPU-mode and employ OpenMP parallelization, the VexCL back-end is 1,5x faster then the slowest one for the largest problem size, which consumes 1,25 GB of main memory.

The same mini-app has been run on an IBM Power S822LC server, which consists of 128 cores running at 4.02 GHz and features Nvidia's NVLink interconnect to communicate with four Nvidia P100 Pascal GPUs. The results are given in columns 5--7 of Table~\ref{tab:mini-app}. The performance of the two CPU back-ends surprisingly differs by 5x. The savings in terms of computing times resulting from using a single GPU is up to 30x.

\begin{table}[h!]
\caption{Wall-clock times (in $\mu s$) measured for the mini-app on different hardware platforms.}
\begin{center}
\begin{tabular}{r|rrr|rrr|}
\hline
        & \multicolumn{3}{c|}{2x Intel E5-2670 @ 2.6 GHz}
        &\multicolumn{3}{c|}{POWER8NVL @ 4.02 GHz + GP100GL}
        \\
\hline
Problem size & Blaze & Eigen & VexCL & Blaze & Eigen & CUDA-VexCL\\
\hline
1.024 & 428	& 420 & 575 & 367 & 149 & 4.487\\
\hline
2.048 & 1.264 & 1.282 & 1.385 & 743 & 322 & 1.076\\
\hline
4.096 & 2.667 & 2.810 & 2.383 & 1.476 & 640 & 1.074\\
\hline
8.192 & 5.100 & 5.330 & 4.122 & 2.941 & 1.277 & 1.158\\
\hline
16.384 & 9.610 & 9.286 & 7.134 & 5.932 & 2.606 & 5.292\\
\hline
32.768 & 17.390 & 15.981 & 12.135 & 11.857 & 5.095 & 2.896\\
\hline
65.536 & 29.907 & 25.778 & 19.024 & 353.000 & 10.495 & 4.644\\
\hline
131.072 & 52.377 & 49.042 & 34.706 & 332.000 & 22.113 & 10.111\\
\hline
262.144 & 110.000 & 110.000 & 71.622 & 126.000 & 46.626 & 16.983\\
\hline
524.288 & 250.000 & 197.000 & 126.000 & 48.985 & 93.918 & 32.395\\
\hline
1.048.576 & 338.000 & 325.000 & 214.000 & 68.930 & 189.000 & 33.759\\
\hline
2.097.152 & 605.000 & 588.000 & 398.000 & 104.000 & 378.000 & 36.770\\
\hline
4.194.304 & 1.119.000 & 1.086.000 & 750.000 & 211.000 & 771.000 & 49.715\\
\hline
8.388.608 & 2.166.000 & 2.114.000 & 1.442.000 & 321.000 & 1.444.000 & 77.291\\
\hline
16.777.216 & 4.222.000 & 4.131.000 & 2.862.000 & 639.000 & 2.910.000 & 128.000\\
\hline
33.554.432 & --- & --- & --- & 1.244.000 & 5.775.000 & 224.000\\
\hline
67.108.864 & --- & --- & --- & 2.353.000 & 11.660.000 & 382.000\\
\hline
134.217.728 & --- & --- & --- & 4.634.000 & 23.692.000 & ---\\
\hline
268.435.456 & --- & --- & --- & 9.683.000 & 47.941.000 & ---\\
\hline
\end{tabular}
\end{center}
\label{tab:mini-app}
\end{table}

\section{CONCLUSIONS}
\label{sec:conclusions}

This paper described the main features and design principles of the FDBB project (\url{https://gitlab.com/mmoelle1/FDBB}) and provided a brief performance analysis. From the results obtained from a synthetic micro-benchmark we conclude that the additional abstraction layer introduced by FDBB does not cause any performance penalty. Preliminary timings for the more realistic mini-app support our claim that it is possible to write generic codes for heterogeneous HPC systems without sacrificing efficiency. It is, however, necessary to implement mechanisms to choose the optimal back-end for the platform at hand since the performance of the different back-ends can differ by orders of magnitudes.

\section*{ACKNOWLEDGEMENTS}
The author would like to thank Denis Demidov, Peter Gottschling, Klaus Iglberger, Karl Rupp, and Conrad Sanderson for their support on integrating the different linear algebra back-ends into FDBB. This project has received funding from the European Union’s Horizon 2020 research and innovation programme under grant agreement No 678727.

\end{document}